\documentclass[preprint]{revtex4}

\usepackage{plain}    % for eqalign
\usepackage{graphicx}
\usepackage{bm}       % bold math
\usepackage{mathptmx} % postscript fonts

\long \def \blockcomment #1\endcomment{}

\begin{document}

\title{The Standard Model and the Lattice}
\author{Michael Creutz}
\affiliation{
  Senior Physicist Emeritus\\
  Physics Department\\
  Brookhaven National Laboratory\\
Upton, NY 11973, USA
}

\email{mike@latticeguy.net}

\begin{abstract}
  { The $SU(3)\otimes SU(2) \otimes U(1)$ standard model maps smoothly
    onto a conventional lattice gauge formulation, including the
    parity violation of the weak interactions.  The formulation makes
    use of the pseudo-reality of the weak group and requires the
    inclusion a full generation of both leptons and quarks.  As in
    continuum discussions, chiral eigenstates of the Dirac operator
    generate known anomalies, although with rough gauge configurations
    these are no longer exact zero modes of the Dirac operator.  }
\end{abstract}

\maketitle

Lattice gauge theory has a long history of successes in the study of
low energy QCD, the underlying theory of the strong nuclear force.
One might ask if the approach could also be used for the weak
interactions related to beta decay.  This is probably of little value
for calculations since the electroweak coupling is small and
conventional perturbation theory is highly accurate for most purposes.
However, from a theoretical point of view a lattice theory provides a
path towards a mathematical definition of a field theory in the limit
where the lattice spacing is taken to zero.  Putting the weak
interactions on the lattice is a first step toward a formal definition
of the theory.  Within the picture, gauge field topology should give
rise to known anomalies, including the weak interactions generating
baryon decay through the effective vertex elucidated by 't Hooft
\cite{'tHooft:1976fv}.  A crucial ingredient is the need to include
entire generations to properly cancel anomalies.  The approach is
similar in spirit to
Refs. \cite{Swift:1984dg,Smit:1985nu,Shrock:1985un,Lee:1986kc}, with
the main difference being the necessity to mingle both the weak and
strong groups.  Although the approach can account for the parity
violation in weak decays, the inclusion of electromagnetism and a
non-trivial Higgs potential require non-asymptotically free couplings.
Thus the path to a rigorous definition of the continuum limit remains
elusive.

To remain as close to traditional lattice methods
\cite{Wilson:1974sk,Wilson:1975id,creutz_2023} as possible, consider
all gauge fields as unitary group elements on the bonds of a four
dimensional hypercubic lattice.  This includes elements of $SU(3)$ for
the strong interactions, $SU(2)$ for the weak isospin group and $U(1)$
for hyper-charge, reflecting the fields of the usual standard model
\cite{Oerter:2006iy}.  Denote these bond variables correspondingly as
$U_{su3},U_{su2}, U_y$.  These fields self-interact through the
standard gauge invariant plaquette form, although nothing here
precludes improvement schemes.  The goal is to maintain exact local
gauge symmetry under all three groups.

For one generation, include eight fermion fields, represented by
\begin{equation}
  u^r,u^g, u^b, d^r,d^b,d^g,\nu,e^-.
  \label{fermions}
  \end{equation}
The three colors $\{r,g,b\}$ are explicit for the up and down quarks
$u,d$. Included also are the neutrino field $\nu$ and the electron
$e^-$.  These are all anticommuting Grassmann variables located on the
lattice sites.  In addition on each site are independent conjugate
Grassmann variables
\begin{equation}
  \overline {u^r},\overline {u^g},\overline { u^b},
  \overline { d^r},\overline {d^b},\overline {d^g},\overline \nu,
  \overline{e^-}.
\end{equation}
With more generations this pattern is repeated for each.

All fermion fields are four component Dirac spinors.  These can be
divided into right and left handed parts
\begin{eqnarray}
  \psi_L=(1-\gamma_5)\psi/2\cr
  \psi_R=(1+\gamma_5)\psi/2\cr
  \overline\psi_L=\overline\psi(1+\gamma_5)/2\cr
  \overline\psi_R=\overline\psi(1-\gamma_5)/2.\cr
\end{eqnarray}
The weak interactions only couple directly to the left handed parts
while the strong group and hyper-charge see both chiralities.  Within
the single generation of fermions, the three gauge groups behave
rather differently.  For the strong interactions there are two
vector-like triplets under $SU(3)$
\begin{equation}
u=\left(
\matrix{u^r \cr u^g \cr u^b\cr}
\right),
\qquad 
d=\left(
\matrix{d^r \cr d^g \cr d^b\cr}
\right).
\end{equation}
In contrast, under the weak interactions there are four left handed
doublets
\begin{equation}
r=\left(\matrix{u^r\cr d^r\cr}\right)_L,\qquad 
g=\left(\matrix{u^g\cr d^g\cr}\right)_L,\qquad
b=\left(\matrix{u^b\cr d^b\cr}\right)_L,\qquad 
l=\left(\matrix{\nu\cr e^-\cr}\right)_L. 
\end{equation}
Here the doublets are labeled by their color or lepton nature.
Finally hyper-charges for the 8 fermions are taken as conventional in
the standard model, differing between the left and right handed
fermions.  As listed in Eq. (\ref {fermions}) for the left handed
parts the assignments are
\begin{equation}
Y_L=(1/3,1/3,1/3,1/3,1/3,1/3,-1,-1),
\end{equation}
and for the right handed components
\begin{equation}
  Y_R=(4/3,4/3,4/3,-2/3,-2/3,-2/3,0,-2).
\end{equation}
For the right handed fields these values are twice the electromagnetic
charge while for the left handed parts they differ from twice the
physical charges by $\pm 1$.  All gauge fields are neutral under
hyper-charge.

The local gauge symmetries correspond to rotations of the 
various fields on the lattice sites.  The strong group acts on
the two quark triplets
\begin{equation}
\psi_{ud}\rightarrow g_{su3} \psi_{ud}.
\end{equation}
Meanwhile the weak group acts on left handed doublets but leaves
the right hand fields untouched
\begin{equation}
\psi_{rgbl}\rightarrow  \left( g_{su2} {1-\gamma_5\over 2} 
+  {1+\gamma_5\over 2}\right)\psi_{rgbl}.
\end{equation}
In addition to the fermions, on each site
is a complex doublet Higgs field
\begin{equation}
  H=\pmatrix{H_1\cr H_2}.
\end{equation}
Both components of this field have hypercharge $Y=1$.
This field also rotates under the weak isospin
\begin{equation}
  H\rightarrow g_{su2}H
\end{equation}
but does not see the strong group.

The group $SU(2)$ is pseudo-real in the sense that a unitary
transformation relates the complex conjugate of any element to itself.
With the usual conventions
\begin{equation}
  g^*=\tau_2 g \tau_2.
\end{equation}
This means that in addition to the initial Higgs doublet there is
another combination
\begin{equation}
  H^\prime\equiv\tau_2 H^* \tau_2=
  \left(\matrix{-H_2^*\cr H_1^*}\right)
  \end{equation}
that transforms equivalently
\begin{equation}
  H^\prime\rightarrow g_{su2}H^\prime
\end{equation}
with hyper-charge $Y=-1$.  Finally the hyper-charge gauge group
rotates the phases of all fields by an angle proportional to their
respective hyper-charges
\begin{equation}  \psi\rightarrow e^{i\theta Y_\psi}\psi.
\end{equation}

It is important that these three gauge groups commute with each other.
The weak group does not change the colors of the quarks.  The strong
group doesn't break weak isospin.  And the hyper-charge assignments
are constant within each chirality of the strong and weak multiplets.

The appearance of an even number of fundamental weak doublets is
essential.  Witten \cite{Witten:1982fp} has discussed how a closed
path in $SU(2)$ field space can change the sign of the fermion
determinant.  This phase ambiguity is extensively reviewed in
\cite{Poppitz:2010at} and is closely related to the 't Hooft vertex
\cite{'tHooft:1976fv} connection to spin flips from the anomaly in
vector multiplets.  With only a left hand multiplet there is nothing
to flip into.  With an even number of multiplets this can be
compensated among them, as discussed later.

The couplings of the fields to the various gauge bosons are contained
in the standard hopping terms between sites, including such terms for
the Higgs field.  All left handed weak doublets rotate by the
$U_{su2}$ while the right handed counterparts do not see that matrix.
The leptons do not interact with the $U_{su3}$ while the quarks do.
All fermions rotate appropriately under hyper-charge.

To complete the picture requires the Higgs mechanism
\cite{Higgs:1964pj,Weinberg:1967tq,Salam:1968rm,Englert:1964et,Guralnik:1964eu}.
This will serve two purposes, generating masses and eliminating
doublers associated with naive fermion hopping.  The Higgs mass is
adjusted via a self interaction potential $V(|H|)$, as in the
continuum treatment of the standard model.  A quartic term is
necessary in $V$ in order to adjust the expectation value $v=\langle
|H|\rangle$ to map directly onto conventional continuum phenomenology.

Concentrating on the weak interactions, for each fermion doublet there
are two distinct on site combinations that are invariant under the
weak gauge group
\begin{equation}
  \matrix{ H^\dagger\psi=H_1^*\psi_1+H_2^*\psi_2\cr
    {H^\prime}^\dagger\psi=-H_2\psi_1+H_1\psi_2.\cr}
  \end{equation}
It is convenient to divide out the Higgs expectation $v$ and think of
the physical left handed particles as ``composite'' with physical
charges $Q$
\begin{equation}
\matrix{
  &e_L&=&H^\dagger l/v             && Q=(Y_l-Y_H)/2=-1\cr
&\nu_L&=&{H^\prime}^\dagger l/v   && Q=(Y_l+Y_H)/2=0\cr
&{u_{[rgb]}}_L&=&H^\dagger{[rgb]}/v && Q=(Y_{rgb}-Y_H)/2=2/3\cr
&{d_{[rgb]}}_L&=&{H^\prime}^\dagger{[rgb]}/v  && Q=(Y_{rgb}+Y_H)/2=-1/3.\cr
}
\end{equation}
This is similar to working perturbatively in the ``unitary'' gauge,
although in proper lattice gauge spirit all gauges are integrated
over.  Indeed, without gauge fixing, the Higgs field rotates rapidly
over directions and cannot have an expectation value.

For each left handed doublet one can form a combination which is
invariant under the weak group
\begin{equation}
  \chi_L={1\over v}\pmatrix{H^\dagger \psi_L\cr {H^\prime}^\dagger \psi_L\cr}
\end{equation}
This allows construction of gauge singlet mass terms for the doublet
\begin{equation}
  \overline\psi_R M  \chi_L + h.c.
\end{equation}
Here $M$ is an arbitrary mass matrix and can include any
inter-generational mixings.  To be consistent with the strong group,
$M$ should be independent of color.

The same mechanism that gives the fermions masses can now be adapted
to eliminate doublers using a Wilson \cite{Wilson:1975id} like
mechanism.  First remove the $SU(2)$ dependence with the Higgs field
as in the mass term.  Then include the Wilson projection operator
$(1\pm\gamma_\mu)/2$ for fermions to hop to neighboring sites.  Thus,
for each doublet hopping from site $i$ to $i+e_\mu$ add to the action
\begin{equation}
  {\overline\psi_R}_{i+e_\mu}(1+\gamma_\mu){\chi_L}_i/2
  +{\overline\psi_R}_{i}(1-\gamma_\mu){\chi_L}_{i+e_\mu}/2
  + h.c. 
\end{equation}
Here the appropriate strong and hyper-charge matrices are suppressed
for notational simplicity.  In a sense this term mimics an irrelevant
operator proportional to $\overline\psi\partial^2\psi$ which moves all
doubler masses to the cutoff scale.  As with the usual Wilson
procedure, this requires an additive mass renormalization.  Because of
this all masses need to be fine tuned.  In this approach the smallness
of neutrino masses is not natural.

Combining the Higgs field with the $SU(2)$ bond variables allows
construction of gauge invariant operators to create the physical W
and Z bosons.  For example
\begin{equation}
  W^+_\mu\sim {H^\prime}^\dagger_{i+e_\mu}\  {U_{su2}}_{i+e_\mu,i}\ H_i
\end{equation}
has charge $Q=1$ and represents the $W^+$ with associated spin in the
bond direction.  Similarly the $W^-$ corresponds to
\begin{equation}
W^-\sim H_{i+e_\mu}^\dagger\ {U_{su2}}_{i+e_\mu,i}\ H^\prime_i.
\end{equation}
In addition there are two neutral combinations
\begin{equation}
\matrix {{H^\prime}^\dagger_{i+e_\mu}\ {U_{su2}}_{i+e_\mu,i}\ H^\prime_i\cr
                 H^\dagger_{i+e_\mu}\ {U_{su2}}_{i+e_\mu,i}\ H_i.\cr }
\end{equation}
While these generally mix, the second term appears in the action as
the hopping term for the Higgs field.  Remaining is an operator for
the physical $Z$.

Note that the parameter count is essentially the same as in the usual
continuum discussions of the standard model.  In addition to the three
independent gauge couplings and the Higgs parameters, all fermion
masses need to be tuned to their physical values. In particular, an
explanation for the small neutrino masses is lacking.  The Wilson
parameter represents the cutoff scale for the fermions and as usual
should be connected to the lattice spacing.

This basically completes the model, but it is instructive to consider
how the usual quantum anomalies come into play.  With dimensional
regularization these effects appear via the fermionic measure not
being chirally symmetric \cite{Fujikawa:1979ay}.  With Wilson
fermions, the anomalies are moved into the behavior of heavy doubler
states.  This is similar in spirit to discussions of anomalies
with Pauli-Villars regulation
\cite{Pauli:1949zm,Adler:1969er,Bell:1969ts} with additional heavy
states added near the cutoff.

In continuum discussions anomalies are frequently tied to topology in the
gauge fields and the index theorem \cite{Atiyah:1963zz,Atiyah:1971rm}
relating to zero modes in the Dirac operator.  On the lattice the
space of allowed configurations is simply connected and does not
support separate topological sectors absent some sort of smoothing
condition \cite{Luscher:1981zq}.  However such restrictions destroy
reflection positivity \cite{Creutz:2004ir} and will interfere with any
Hamiltonian formulation.

It is interesting to contrast this picture with the overlap approach
of Neuberger \cite{Neuberger:1997fp,Narayanan:1992wx,Narayanan:1993sk}
where one projects the relevant eigenvalues onto exact zero modes.
This has been successful to all orders in peerturbation theory
\cite{Luscher:2000zd}.  It does, however, eliminate exact locality of
the Dirac operator \cite{Horvath:1998cm,Horvath:1999bk}.  In addition
the projection process encounters singularities as one transits
between topological sectors \cite{Creutz:2002qa}.  In the current
approach, the Dirac operator remains local while robust zero modes are
lost.

The formulation presented here preserves gamma-five hermeticity for
the Dirac operator
\begin{equation}
\gamma_5 D \gamma_5= D^\dagger.
\end{equation}
Indeed most lattice fermion prescriptions, with the exception of
twisted mass \cite{Frezzotti:2000nk}, satisfy this.  An immediate
consequence is that on diagonalizing $D$ all eigenvalues are either
real or in complex conjugate pairs.  (Since D is not a
normal operator, consider either left or right eigenvalues for this
discussion.)

\begin{figure*}
\centering
\includegraphics[width=3in]{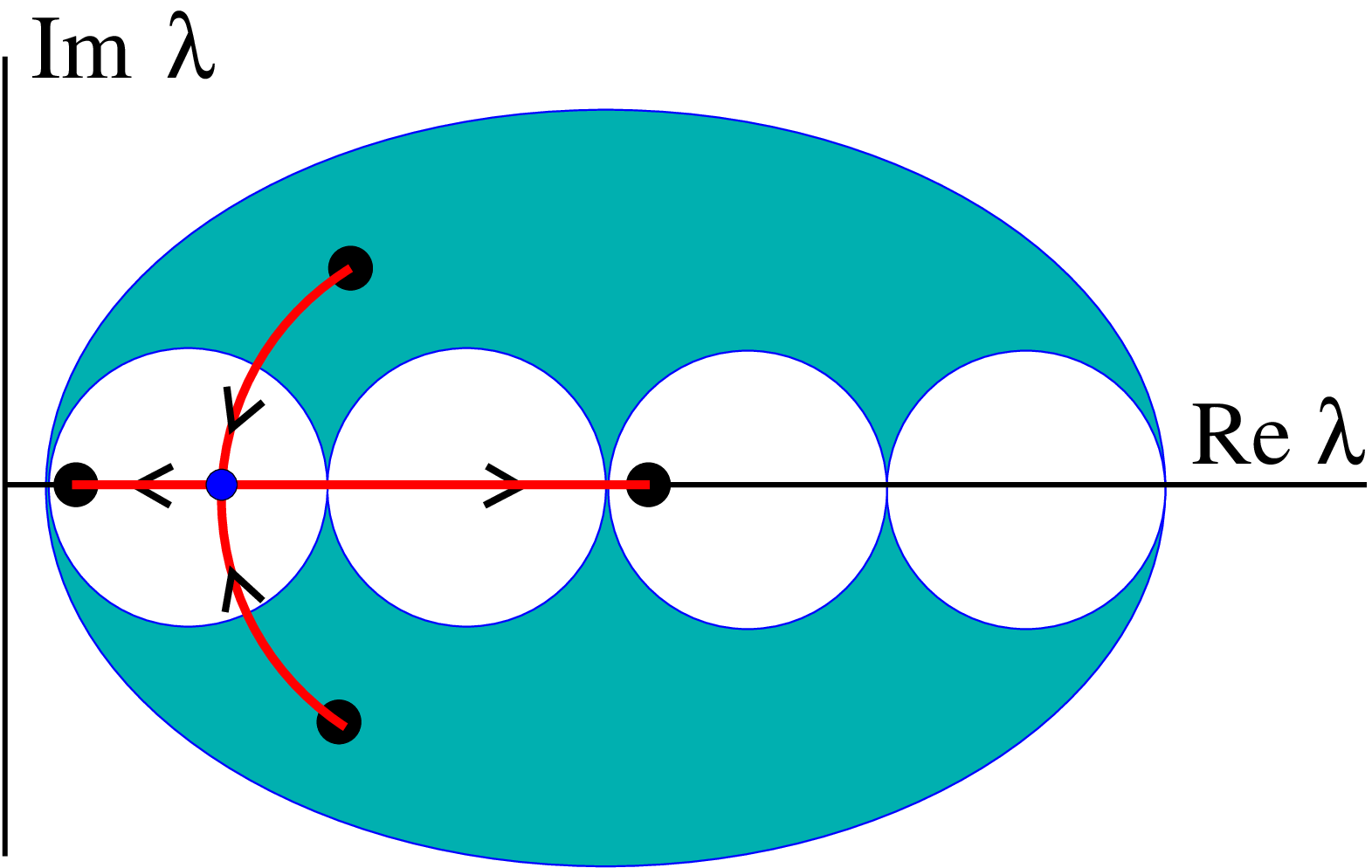}
\caption{\label{eigenflow}
\sl
  The merging of two complex eigenvalues to
  form two real ones, one of which can become small with the other
  moving into the doubler region.  The background region represents
  the spectrum of free Wilson fermions.  If the fields are smoothed
  into a classical instanton, the small eigenvalue becomes the zero
  mode from topology and the index theorem.  }
\end{figure*}

If the gauge fields are sufficiently smooth, the index theorem does
apply and modes of non-trivial chirality are well known.  However,
since the space of lattice fields is simply connected, there must
exist a path connecting a configuration with such chiral states to one
without, as discussed in Ref. \cite{Creutz:2002qa}.  In terms of the
eigenvalues of $D$, a complex conjugate pair joins on the real axis
and splits apart as two real eigenvalues.  One can move to small real
part while the other moves off to the doubler region, as sketched in
Fig. \ref{eigenflow}.  Ref. \cite{Creutz:2010ec} demonstrated that
such a path does does exist and does not need to pass through a
barrier of of large action, although it does require local fields to
violate the smoothness condition of \cite{Luscher:1981zq}.

Consider the space spanned by the real eigenvalues of $D$.  On this
sub-space $\gamma_5$ commutes with $D$ and can be simultaneously
diagonalized.  Thus the states can be labeled by chirality.  The usual
topological structures are represented by an imbalance of small
eigenvalues of one chirality over the other.  If a gauge field
configuration with such a mode is now smoothed, the small eigenvalue
will be driven to zero and satisfy the continuum index theorem.  As
the full trace of $\gamma_5$ must vanish, such zero modes have
corresponding modes of the opposite chirality in the doubler region.

These chiral eignmodes are directly tied to quantum anomalies as
discussed by 't Hooft \cite{'tHooft:1976fv}.  Small or zero eigenvalues
suppress the partition function
\begin{equation}
Z=\int (dA)\ (d\overline \psi d\psi)
e^{-S_g+\overline\psi D \psi}
=\int (dA)\ e^{-S_g(A)}\ \prod \lambda_i. 
\end{equation}
On the surface, this suggests that zero modes are irrelevant as they
don't appear in the partition function.  But 't Hooft showed how
certain observables can overcome this suppression.  To see this, first
introduce abstract sources $\eta$ and $\overline \eta$
\begin{equation}
 Z(\eta,\overline\eta)=\int (dA)\ (d\overline \psi d\psi)\ 
e^{-S_g+\overline\psi D \psi +\overline\psi \eta+
  \overline\eta\psi}.
\end{equation}
Differentiation (in a Grassmann sense) with respect to the sources
generates the Green's functions of the theory.  Completing the square
and doing the integral over the fermions gives
  \begin{equation}
 Z=\int (dA)\ 
e^{-S_g{+\overline\eta D^{-1} \eta}/4}\ 
\prod \lambda_i.
\end{equation}
If the sources overlap with one of the small real eigenmodes, the
inverse of the corresponding eigenvalue can enter the Green's function
and cancel the suppression in the partition function.

\begin{figure*}
\centering
\includegraphics[width=3in]{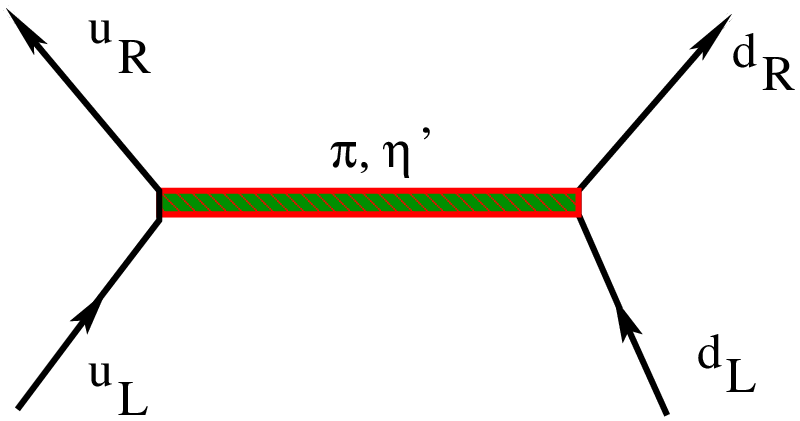}
\caption{\sl \label{eigen1} The chiral anomaly produces a spin flip
  amplitude involving all quark species.  The non-vanishing of this
  diagram induces a mass for the eta prime and causes a mixing of the
  up and down quark masses when they are not degenerate.
  Figure taken from Ref. \cite{Creutz:2010ts}.
}
\label{fourpoint}
\end{figure*}

 This effect is well understood for the strong interactions.  A chiral
 eigenmode couples left handed quarks to right handed ones, resulting
 in a non-vanishing spin flip amplitude, even if the quarks are
 mass-less.  The mixing of various pseudo-scalars through the process
 is sketched in Fig. \ref{fourpoint}.  This vertex is tied to the mass
 of the eta-prime meson.  Except in certain special cases, such as the
 large number of colors limit \cite{Witten:1979vv}, the resulting
 value needs to be determined through simulations.  Note that there
 remains a second chiral eigenmode in the doubler region, but due to
 the large masses of the doublers this only gives a constant factor to
 the partition function.

The consequences of the anomaly for the weak interactions are less
familiar, primarily since the effect is quite small.  The
non-perturbative chiral modes will be suppressed exponentially in the
inverse electroweak coupling.  Although tiny, the effects are crucial
to understanding the structure of the theory.  For each left handed
$SU(2)$ doublet, its conjugate field is right handed
\begin{equation}
\psi^c=\tau_2\gamma_2\psi^*.
\end{equation}
For example, the anti-neutrino is right handed.  Of our four doublets
$\{r,g,b,l\}$, take two and pair them with the conjugates of the other
two and then antisymmetrize over the combinations.  Using sources that
overlap with a low chiral mode gives a non-vanishing value for the
four point ``vertex''
\begin{equation}
\epsilon_{ijkl}
\langle \overline \psi_i
^c \overline \psi_j^c
D^{-1} \psi_k \psi_l\rangle \ne 0.
\end{equation}
Here the indices run over the four doublets $\{r,g,b,l\}$.

This effective interaction violates both baryon and lepton number, but
preserves the difference $B-L$.  While this discussion is for the
euclidean formulation of the theory, in a Hamiltonian approach the
process proceeds from modes crossing in and out of the Dirac sea
\cite{Creutz:1993vv,Creutz:2001wp}.  In the process, fermion number
changes by 2 units.  This is consistent with $SU(2)$ since the group
is pseudo-real.  It is also consistent with the $SU(3)$ symmetry since
$\overline 3 \in 3\otimes 3$ and two flavors are going to one
anti-flavor.  A version of this vertex appears in a proposed domain
wall approach to the weak interactions
\cite{Creutz:1996xc,Creutz:1997fv,Creutz:2018dgh}.  The net process
can be thought of as an effective mixing of the anti-neutron with the
neutrino and the anti-proton with the electron
\begin{equation}
\pmatrix{n^c\cr  p^c}_R\Longleftrightarrow \pmatrix {\nu \cr e^-}_L.
\end{equation}  
Through this mechanism proton decay $ p\rightarrow e^+ + \pi$ is
allowed, although it is extremely small being suppressed exponentially
in $1/\alpha$.  As with the eta-prime mass, the numerical value for
this process is dynamical and can only be determined through
simulations.

In summary, one generation of the standard model fits nicely into a
conventional lattice gauge framework. The approach keeps the
$SU(3)\otimes SU(2)\otimes U(1)$ gauge symmetries exact.  To be
consistent with known anomalies, the approach requires including each
generation in its entirety.  The small baryon and lepton violation
demonstrated by 't Hooft appears in the lattice approach through the
appearance of chiral eigenmodes of the Dirac operator.  Unlike in the
continuum where differentiable fields are assumed, these modes are not
forced to occur exactly at zero.  The overall picture is close in
spirit to confining models
\cite{Osterwalder:1977pc,Fradkin:1978dv,Greensite:2017ajx}.  The main
remaining issue concerns asymptotic freedom, absent both in
electromagnetism and the Higgs quartic self coupling.  This leaves an
obstacle towards defining the theory in the continuum limit.  For
electromagnetism this suggests a possible unification with further
gauge fields at high energies.  For the Higgs it hints at composite
models or possibly involving gravity at the highest
energies \cite{Shaposhnikov:2009pv,Alekhin:2012py}.

\section*{Acknowledgments}
This manuscript has been authored under contract number
DE-AC02-98CH10886 with the U.S.~Department of Energy.  Accordingly,
the U.S. Government retains a non-exclusive, royalty-free license to
publish or reproduce the published form of this contribution, or allow
others to do so, for U.S.~Government purposes.

%\bibliography{../references,../creutz}
%\bibliographystyle{unsrt}
%\bibliographystyle{apsrev}

\end{document}